\documentclass[fleqn,usenatbib,useAMS]{mnras}

\usepackage{newtxtext,newtxmath}
\usepackage{braket}
\usepackage{natbib}
\usepackage{booktabs}
\usepackage{multirow}
\usepackage{ulem}
\usepackage[dvipsnames]{xcolor}

\usepackage[T1]{fontenc}

\DeclareRobustCommand{\VAN}[3]{#2}
\let\VANthebibliography\thebibliography
\def\thebibliography{\DeclareRobustCommand{\VAN}[3]{##3}\VANthebibliography}

\usepackage{graphicx}	
\usepackage{amsmath}	

\title[New insights into the He star AM CVn formation]{New insights into the helium star formation channel of AM CVn systems with explanations of Gaia14aae and ZTFJ1637+49}
\author[Sarkar, Ge \& Tout]{
Arnab Sarkar$^{1}$\thanks{E-mail: as3158@cam.ac.uk},
Hongwei Ge$^{2,1,3}$\thanks{E-mail: gehw@ynao.ac.cn},
and Christopher A. Tout$^{1}$\thanks{E-mail: cat@ast.cam.ac.uk}
\\
\\
$^{1}$Institute of Astronomy, The Observatories, Madingley Road, Cambridge CB3 OHA, UK
\\
$^{2}$Yunnan Observatories, Chinese Academy of Sciences, Kunming 650216, China
\\$^{3}$ Key Laboratory for Structure and Evolution of Celestial Objects, Chinese Academy of Sciences, P.O. Box 110, Kunming 650216, China
}

\date{Accepted XXX. Received YYY; in original form ZZZ}

\pubyear{2015}

\begin{document}
\label{firstpage}
\pagerange{\pageref{firstpage}--\pageref{lastpage}}
\maketitle

\begin{abstract}
We model helium-rich stars with solar metallicity ($X=0.7,\:Z=0.02$) progenitors that evolve to form AM Canum Venaticorum systems through a helium-star formation channel, with the aim to explain the observed properties of Gaia14aae and ZTFJ1637+49. We show that semi-degenerate, H-exhausted ($X\leq 10^{-5}$), He-rich ($Y\approx0.98$) donors can be formed after a common envelope evolution (CEE) phase if \textcolor{black}{either additional sources of energy are used to eject the common envelope, or a different formalism of CEE is implemented.} We follow the evolution of such binary systems after the CEE phase using the Cambridge stellar evolution code, when they consist of a He-star and a white dwarf accretor, and report that the mass, radius, and mass-transfer rate of the donor, the orbital period of the system, and the lack of hydrogen in the spectrum of Gaia14aae and ZTFJ1637+49 match well with our modelled trajectories wherein, after the CEE phase Roche lobe overflow is governed not only by the angular momentum loss (AML) owing to gravitational wave radiation ($\mathrm{AML_{GR}}$) but also an additional AML owing to $\alpha-\Omega$ dynamos in the donor. This additional AML is modelled with our double-dynamo (DD) model of magnetic braking in the donor star. We explain that this additional AML is just a consequence of extending the DD model from canonical cataclysmic variable donors to evolved donors. We show that none of our modelled trajectories match with Gaia14aae or ZTFJ1637+49 if the systems are modelled only with $\mathrm{AML_{GR}}$.

\end{abstract}


\section{Introduction} 
AM Canum Venaticorum (AM CVn) stars are a class of semi-detached binaries with extremely short orbital periods, $10\lesssim~P_\mathrm{orb}/\,\mathrm{min}\lesssim 65$. Although closely related to cataclysmic variables (CVs), these systems have shorter orbital periods and usually lack H in their spectrum \citep{Solheim2010}. They are usually modelled by an evolved star transferring mass to a white dwarf (WD) accretor. Owing to their short orbital periods, these systems are strong gravitational wave sources \citep{Kupfer2016}. 

Three possible formation channels for AM CVn systems have been \textcolor{black}{proposed}. These differ from each other based on the number of common envelope evolution (CEE) phases the primordial main-sequence (MS) binary goes through and the nature of the donor star. In the first formation channel, known as the WD channel, the donor is a He WD which commences Roche lobe overflow (RLOF) and transfers mass to a more massive \textcolor{black}{carbon-oxygen} (C/O) WD after going through two common envelope (CE) phases (see \citealt{Deloye2007} and references therein). The second channel is known as the He-star channel, wherein the donor commences RLOF as either a non-degenerate or semi-degenerate He-rich or He-burning star and transfers mass to a WD after going through two CE phases (see \citealt{Yungelson2008} and references therein). The final channel is known as the Evolved CV channel \textcolor{black}{in which} an evolved MS star commences stable RLOF after going through a single CE phase and transfers mass to a WD accretor \textcolor{black}{while} in the Hertzprung gap (between the end of its MS and the beginning of its \textcolor{black}{ascent of the} red giant branch,  \citealt{Podsiadlowski2003}). In this work we focus on the He-star channel, and refer the reader to  \cite{Solheim2010} for a thorough discussion of each formation channel.

A major challenge in modelling AM CVn systems through the He-star channel is to obtain the system properties, such as the initial mass of the donor and the orbital separation of the system, of the He-star plus WD binary, just after it emerges from the second CEE which forms the naked He-star. This is because the CE phase is still not well understood, with poor constraints on parameters such as the time-scales and the energies involved (see \citealt{2013A&ARv..21...59I} for a review of the CE evolution). Hereinafter, we shall refer to the second CEE whenever we mention CEE. The He-star channel is also sensitive to the evolutionary state of the donor\footnote{We refer to the star undergoing (stable of unstable) RLOF as the donor.} when dynamically unstable RLOF commences. This can range from core He-rich ($Y_\mathrm{c}\approx0.98$) to core He-exhausted ($Y_\mathrm{c}\approx0$) donors, where $Y_\mathrm{c}$ denotes central He-mass fraction. As a consequence, the corresponding AM CVn spectra may show large variances in their abundance profiles.

Another aspect that has been overlooked till now is the inclusion of additional mechanisms by which angular momentum is lost from these systems. Owing to their short orbital periods, usually less than $60 \,\mathrm{min}$, all previous work has assumed that angular momentum loss (AML) by gravitational wave radiation ($\mathrm{AML_{GR}}$) is the only mechanism which drives the evolution of AM CVn systems (e.g. \citealt{Deloye2007} for the WD channel, \citealt{Yungelson2008} for the He-star channel and \citealt{Podsiadlowski2003} for the Evolved CV channel). Although this assumption is reasonable at $P_\mathrm{orb}\approx 10\,\mathrm{min}$, owing to the strong dependence of $\mathrm{AML_{GR}}$ on the orbital separation (and hence $P_\mathrm{orb}$, we repeat that the observed AM CVn systems span $10\lesssim~P_\mathrm{orb}/\,\mathrm{min}\lesssim 65$), at these periods other AML mechanisms might be at play. This is revealed by the disagreement between the theoretically predicted orbital period minimum ($P_\mathrm{orb,mini}$) spike and observed orbital period minimum spike in the orbital period distribution of CVs \citep{Gnsicke2009}. Theoretical models, solely evolved with $\mathrm{AML_{GR}}$, predict $P_\mathrm{orb,mini}\approx65\,\mathrm{min}$ whereas observations suggest $P_\mathrm{orb,mini}\approx80\,\mathrm{min}$. \cite{Knigge2011} attempt to alleviate this discrepancy by multiplying $\mathrm{AML_{GR}}$ by 2.47. Although this modification leads to better agreement between theory and observations, the multiplicative factor, being ad hoc, is not a reliable probe to use to explain the evolution of AM CVn systems, which involves evolved donors. \citet[hereinafter SGT]{SGT} have included an additional AML mechanism in order to model AM CVn systems through the Evolved CV channel. In this model the AML is driven by two $\alpha-\Omega$ dynamos operating in the donor star. This physically motivated formulation was used by \citet[hereinafter ST]{2022MNRAS.513.4169S} to explain the observed properties of CVs. However, a detailed analysis of how such additional AML mechanisms extend from operating in canonical CVs to operating in AM CVn systems is yet to be made in full. Our reason for revisiting the He-star channel is two peculiar systems, Gaia14aae, reported by \cite{Campbell2015}, with observed properties reported by \cite{Green2018} and \cite{Green2019}, and ZTF1637J+49, reported by \cite{2022MNRAS.512.5440V}. Here we examine the He-star channel of AM CVn formation with the DD model to explain these two systems.

In Section~\ref{sec:moti} we explain why Gaia14aae and ZTFJ1637+49 challenge our usual understanding of AM CVn evolution. In Section~\ref{sec:cee} we show how the CEE can result in viable He-star plus WD candidates which can evolve to explain Gaia14aae and ZTFJ1637+49. In Section~\ref{sec:models}, we use the Cambridge stellar evolution code (STARS) to track the detailed evolution of these candidates. In Section~\ref{sec:dd} we demonstrate the implications of extra AML mechanisms in the analysis of AM CVn systems. We summarize and conclude our work in Section~\ref{sec:conc}.

\section{Motivation}
\label{sec:moti}

Gaia14aae, or ASSASN-14cn, was first identified in ourburst by the All-Sky Automated Survey for Supernovae (ASAS-SN; \citealt{2014ApJ...788...48S}) with follow up observation as a deeply eclipsing, H-deficient system, with $P_\mathrm{orb} = 49.71\,\mathrm{min}$ by \cite{Campbell2015}. \cite{Green2018} showed that it consists of a donor with $M_2 = 0.0250 \pm 0.0013 M_\odot$, $R_2 = 0.0606\pm0.0003R_\odot$, and an accretor of mass $M_1 = 0.87 \pm 0.02 M_\odot$. Upon comparing its orbital properties with the trajectories of the three proposed formation channels, they concluded that Gaia14aae most likely did not descend from the WD or the He-star channel. The trajectories that matched well with its donor properties consisted of systems evolved from H-dominated CVs but such systems ought to show H in their spectra, contradictory to observations. The trajectories shown by SGT (see their figs~14 to 16) concurred with previous studies wherein the observed properties of Gaia14aae matched well with an AM Cvn consisting of a partially H-exhausted donor. \cite{Green2019} detected N and O but not C in the spectrum of Gaia14aae and concluded that Gaia14aae may be an unusual AM CVn descended from the Evolved CV channel. ZTF1637J+49 was one of five detected eclipsing AM CVn systems detected by the Zwicky Transient Facility (ZTF; \citealt{2022MNRAS.512.5440V}). They observe $P_\mathrm{orb} = 61.5\,\mathrm{min}$ and report $M_2 = 0.023\pm0.008M_\odot$, $R_2 = 0.068\pm0.007R_\odot$ and $M_1 = 0.90\pm0.05M_\odot$. Similar to Gaia14aae, no H has been detected in its spectrum. While N was detected in its spectrum there was no detection of C or O. The results of SGT show that ZTF1637J+49 \textcolor{black}{has a higher orbital period and donor radius, and a smaller donor mass than Gaia14aae}, suggesting an even less H-deficient donor star, again contradicting observations. We tabulate the observed parameters we work with in Table \ref{tab:table}\footnote{In this Table, $M_2$ is equivalent to $M_\mathrm{He}$.}.
\begin{table*}
	\centering
	\caption{Binary parameters and detected elements of Gaia14aae and ZTF1637J+49 from \protect\cite{Green2018,Green2019} and \protect\cite{2022MNRAS.512.5440V} respectively. Both are based on modelling the $g$-band lightcurves. Gaia14aae has a mass-loss rate reported by \protect\cite{2018A&A...620A.141R}, while Gaia14aae* has a mass-loss rate reported by \protect\cite{Campbell2015}.}
	\label{tab:table}
	\begin{tabular}{lcccccr} 
		\hline
		 System &$P_\mathrm{orb}/\,\mathrm{min}$ & $M_1/\,M_\odot$ & $M_2/\,M_\odot$ & $R_2/\,R_\odot$& $\Dot{M}_2/\,M_\odot\,\mathrm{yr^{-1}}$& CNO elements detected\\
		\hline
		Gaia14aae & 49.7 & $0.872\pm0.007$ & $0.0253\pm0.0007$ & $0.0603\pm0.0003$&$3.3\times10^{-11}\pm4.3\times10^{-12}$ & N, O\\
		Gaia14aae* & 49.7 & $0.872\pm0.007$ & $0.0253\pm0.0007$ & $0.0603\pm0.0003$&$7-8\times10^{-11}$ & N, O\\
		ZTFJ1637+49 & 61.5 & $0.90\pm0.05$ & $0.023\pm0.008$ & $0.068\pm0.007$& - & N\\

		\hline
	\end{tabular}
\end{table*}


\section{The common envelope evolution outcome}
\label{sec:cee}

In this section we explore the various possible donor stars that can form from the CEE phase. \cite{Green2018}, \cite{Green2019} and \cite{2022MNRAS.512.5440V} show that there is no detectable H in the spectra of Gaia14aae and ZTFJ1637+49. \cite{Green2018} state that any H mass fraction $X$ greater than $10^{-5}$ should trigger detectable Balmer emission, whereas \cite{Green2019} state that an upper limit of $10^{-4}$ for H-exhaustion cannot be ruled out either. So we define a H-exhausted core to be the region of a star with $X\leq10^{-5}$. \cite{Green2019} and \cite{2022MNRAS.512.5440V} also find no detectable C in the spectra of Gaia14aae and ZTFJ1637+49 respectively. Because an increased C abundance arises from convective He burning, we deduce that the progenitor of the donor in these systems must have undergone CEE before He ignition. With these constraints on the abundances of H and C, we sought to find configurations which, after CEE, can give us suitable progenitor He-star donors from which Gaia14aae and ZTFJ1637+49 can evolve. 

\subsection{The energy formalism}
We begin with the energy formalism for the prediction of CEE, given by  

\begin{center}
\begin{equation}
\label{eq:cee_e}
\alpha_\mathrm{CE}(E_\mathrm{orb,f} - E_\mathrm{orb,i}) = E_\mathrm{bind},
\end{equation}
\end{center}
where $E_\mathrm{orb,f}$, $E_\mathrm{orb,i}$ and $E_\mathrm{bind}$ are the final and initial orbital energies and the initial binding energy of the envelope and $\alpha_\mathrm{CE}$ is the efficiency of the common envelope ejection (see \citealt{2013A&ARv..21...59I} for a thorough overview). The orbital energies of the cores after entering the CE phase are given by
\begin{center}
\begin{equation}
\label{eq:cee_eorbi}
E_\mathrm{orb,i} = - \frac{GM_\mathrm{He}M_1}{2a_\mathrm{i}}
\end{equation}
\end{center}
and 
\begin{center}
\begin{equation}
\label{eq:cee_eorbf}
E_\mathrm{orb,f} = - \frac{GM_\mathrm{He}M_1}{2a_\mathrm{f}},
\end{equation}
\end{center}
where $G$ is Newton's gravitational constant, $M_\mathrm{He}$ and $M_1$ the masses of the He-core (which eventually forms the He-star) and the WD accretor, and $a_\mathrm{i}$ and $a_\mathrm{f}$ are the initial and final orbital separations. {\textcolor{black}{We note that some groups instead use $E_\mathrm{orb,i} = - {GM_\mathrm{2}M_1}/{2a_\mathrm{i}}$ (see section~3.3 of \citealt{Han1995}, or equation (3.2) of \citealt{ivanova2020}). This definition has the effect of lowering the net energy release $\lvert E_\mathrm{orb,f} - E_\mathrm{orb,i} \rvert$ in orbital contraction because $M_2>M_\mathrm{He}$. This leads to a general overall increase in $\alpha_\mathrm{CE}$s (equations \ref{eq:cee_alpha1} and \ref{eq:cee_alpha2}). }}We find $a_\mathrm{i}$ by assuming that the donor star overfills its Roche lobe, of radius $R_\mathrm{L}$, at an orbital separation of $a_\mathrm{i}$ such that $R_\mathrm{L}=R_2$. So with the relation between $R_\mathrm{L}$ and $a$ \citep{1983ApJ...268..368E}, we have
\begin{center}
\begin{equation}
\label{eq:cee_rl}
a_\mathrm{i} = \frac{R_2}{0.49}\frac{0.6q^{2/3} + \mathrm{ln}(1 + q^{1/3})}{q^{2/3}} \;\;\mathrm{for}\;\;0<q<\infty,
\end{equation}
\end{center}
where $q= M_2/M_1$ and $M_2$ is the total mass of the donor star before CEE. Later in this section we examine the viability of the CEE to produce configurations with a particular $a_\mathrm{f}$ by varying it as a free parameter. 

In order to find $E_\mathrm{bind}$, we follow the method of \cite{Tout1997} in which the binding energy of the (subgiant or giant) donor is estimated to be 
\begin{center}
\begin{equation}
\label{eq:cee_ebind1}
E_\mathrm{bind} = - \frac{GM_\mathrm{2,env}M_2}{\lambda R_2},
\end{equation}
\end{center}
where $M_\mathrm{2,env}=M_2 - M_\mathrm{He}$ is the mass of the envelope of the donor, and $\lambda$ is a dimensionless parameter governing the structure of the donor. For all our calculations we fix $\lambda=0.5$, \textcolor{black}{although previous works have shown that $\lambda$ varies with the evolutionary state of the donor and its mass (see \citealt{2010ApJ...716..114X} and the references therein). This has implications on the efficiency of common envelope ejection (see Section~\ref{subs:candidates})}. However \cite{Iben1993} instead use the total binding energy of the CE, which surrounds $M_\mathrm{He}$ and $M_1$ with a diameter of about $2a_\mathrm{i}$. With this formalism their expression of $E_\mathrm{bind}$ becomes 
\begin{center}
\begin{equation}
\label{eq:cee_ebind2}
E_\mathrm{bind} = - \frac{GM_\mathrm{2,env}(M_1+M_2)}{2a_\mathrm{i}}.
\end{equation}
\end{center}

With equations (\ref{eq:cee_e} to \ref{eq:cee_ebind2}), we can estimate $\alpha_\mathrm{CE}$ for the two expressions of $E_\mathrm{bind}$ above. When we assume no mass loss from $M_2$ before the commencement of CEE,  $\alpha_\mathrm{CE}$ as a function of time and final separation is given by
\begin{center}
\begin{equation}
\label{eq:cee_alpha1}
\alpha_\mathrm{CE_1}(t,a_\mathrm{f}) = \displaystyle \frac{\frac{\displaystyle 2(M_2 - M_\mathrm{He}(t))M_2}{\displaystyle  R_2(t)}}{\frac{\displaystyle M_\mathrm{He}(t)M_1}{\displaystyle 2a_\mathrm{f}} - \frac{\displaystyle M_\mathrm{He}(t)M_1}{\displaystyle 2a_\mathrm{i}(t)}}
\end{equation}
\end{center}
for $E_\mathrm{bind}$ used by \citet[equation~\ref{eq:cee_ebind1}]{Tout1997}, whereas

\begin{center}
\begin{equation}
\label{eq:cee_alpha2}
\alpha_\mathrm{CE_2}(t,a_\mathrm{f}) = \displaystyle \frac{\frac{\displaystyle (M_2 - M_\mathrm{He}(t))(M_1+M_2)}{\displaystyle 2a_\mathrm{i}(t)}}{\frac{\displaystyle M_\mathrm{He}(t)M_1}{\displaystyle 2a_\mathrm{f}} - \frac{\displaystyle M_\mathrm{He}(t)M_1}{\displaystyle 2a_\mathrm{i}(t)}}
\end{equation}
\end{center}
for $E_\mathrm{bind}$ used by \citet[equation~\ref{eq:cee_ebind2}]{Iben1993}. \textcolor{black}{We point out that owing to a different definition of binding energy, the formalism of \citet[equation~\ref{eq:cee_alpha2}]{Iben1993} leads to an artificially smaller $\alpha_\mathrm{CE}$ than the definition of \citet[equation~\ref{eq:cee_alpha1}]{Tout1997}. This is because \cite{Iben1993} assume that the radius of the CE is about $a_\mathrm{i}$, which is much bigger than the radius of the (sub)giant during RLOF. \citet[section~6.10]{Han1995} argue that although the radius of the CE may be around $a_\mathrm{i}$, most of the mass is concentrated around the donor's envelope. This has a general effect of lowering $E_\mathrm{bind}$ in their formalism (see Section~\ref{subs:candidates} for a comparison of the two expressions of $\alpha_\mathrm{CE}$).}

\subsection{The method}
\label{subs:method}
\begin{figure}
\includegraphics[width=0.5\textwidth]{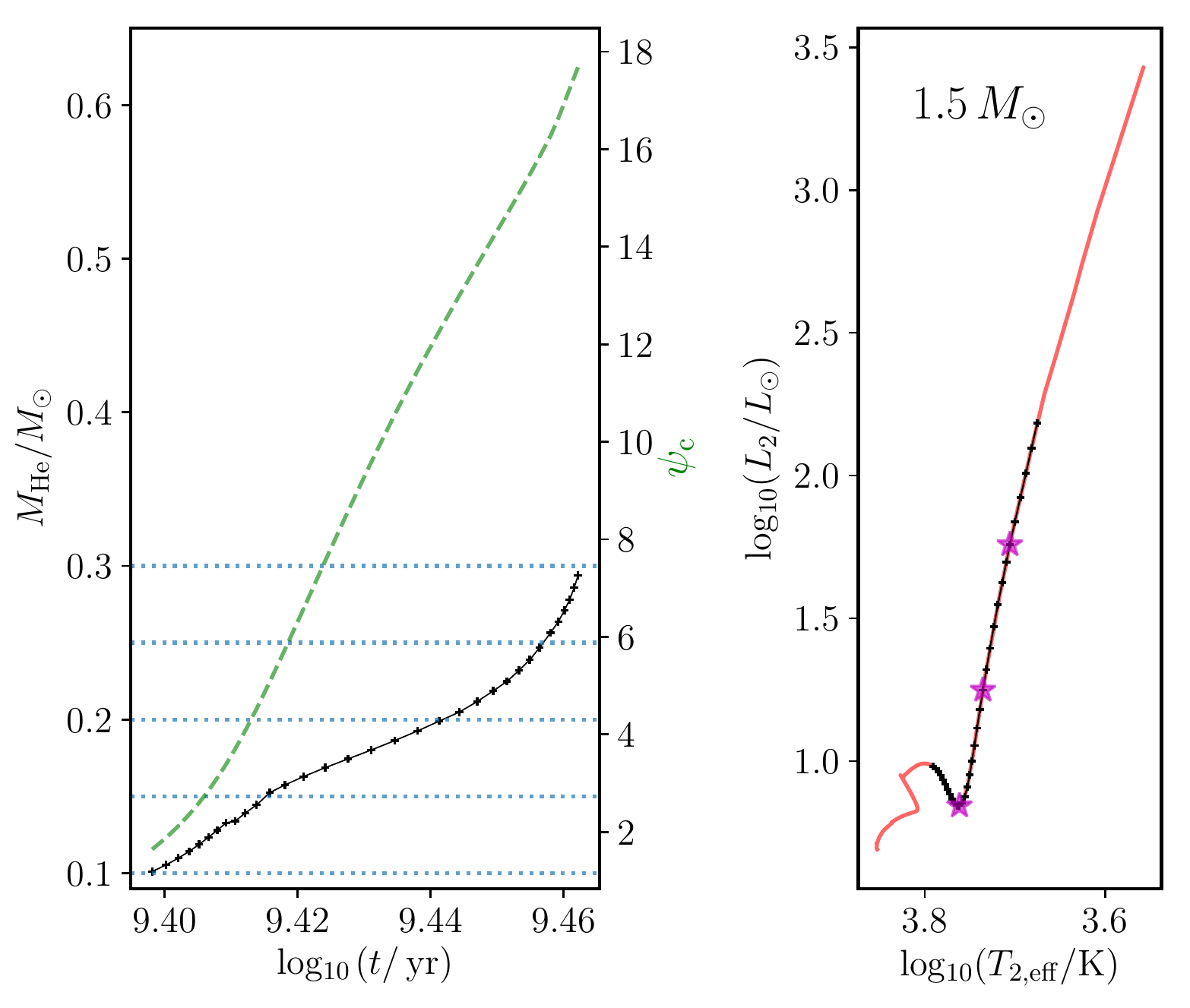}
\caption{The properties of the He core in a $1.5M_\odot$ secondary star. \textit{Left}: The black markers denote the evolution of $M_\mathrm{He}$ with time. The green dashed line shows the time evolution of the central degeneracy parameter $\psi_\mathrm{c}$. \textit{Right}: The HR diagram of these candidates, showing the evolutionary stage of the star when $M_\mathrm{He}$ forms. The blue horizontal dotted lines on the left show $M_\mathrm{He}/M_\odot \in \{0.1,\,0.15,\,0.2,\,0.25,\,0.3\}$ while the stars in magenta on the right show the evolutionary stage of the star when it forms a H-exhausted core of mass $0.15M_\odot$, $0.2M_\odot$ and $0.25M_\odot$.}
\label{fig:ce_1p5}
\end{figure}

\begin{figure}
\includegraphics[width=0.5\textwidth]{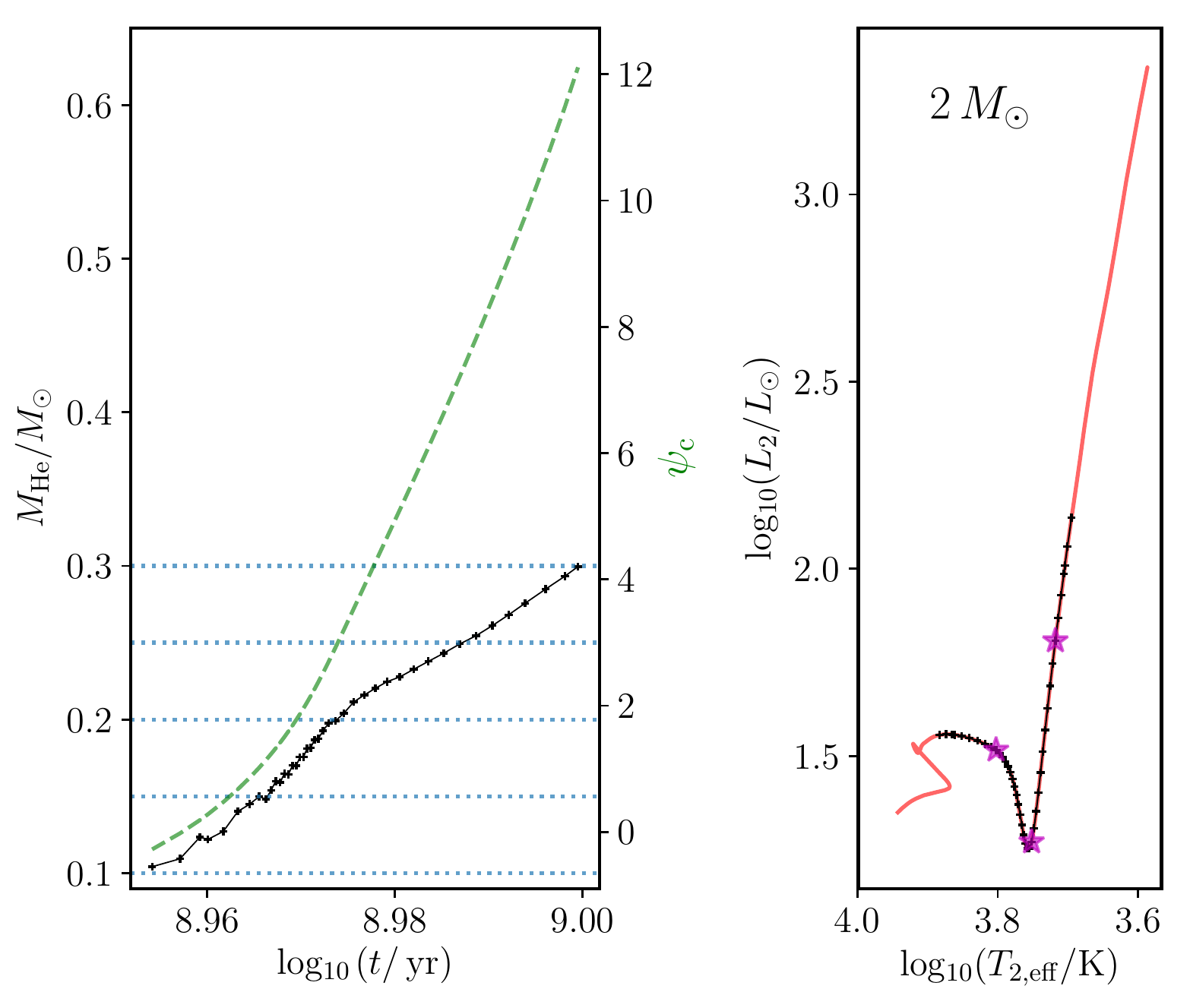}
\caption{Similar to Fig.~\ref{fig:ce_1p5}, but for a $2M_\odot$ secondary star.}
\label{fig:ce_2}
\end{figure}

\begin{figure}
\includegraphics[width=0.5\textwidth]{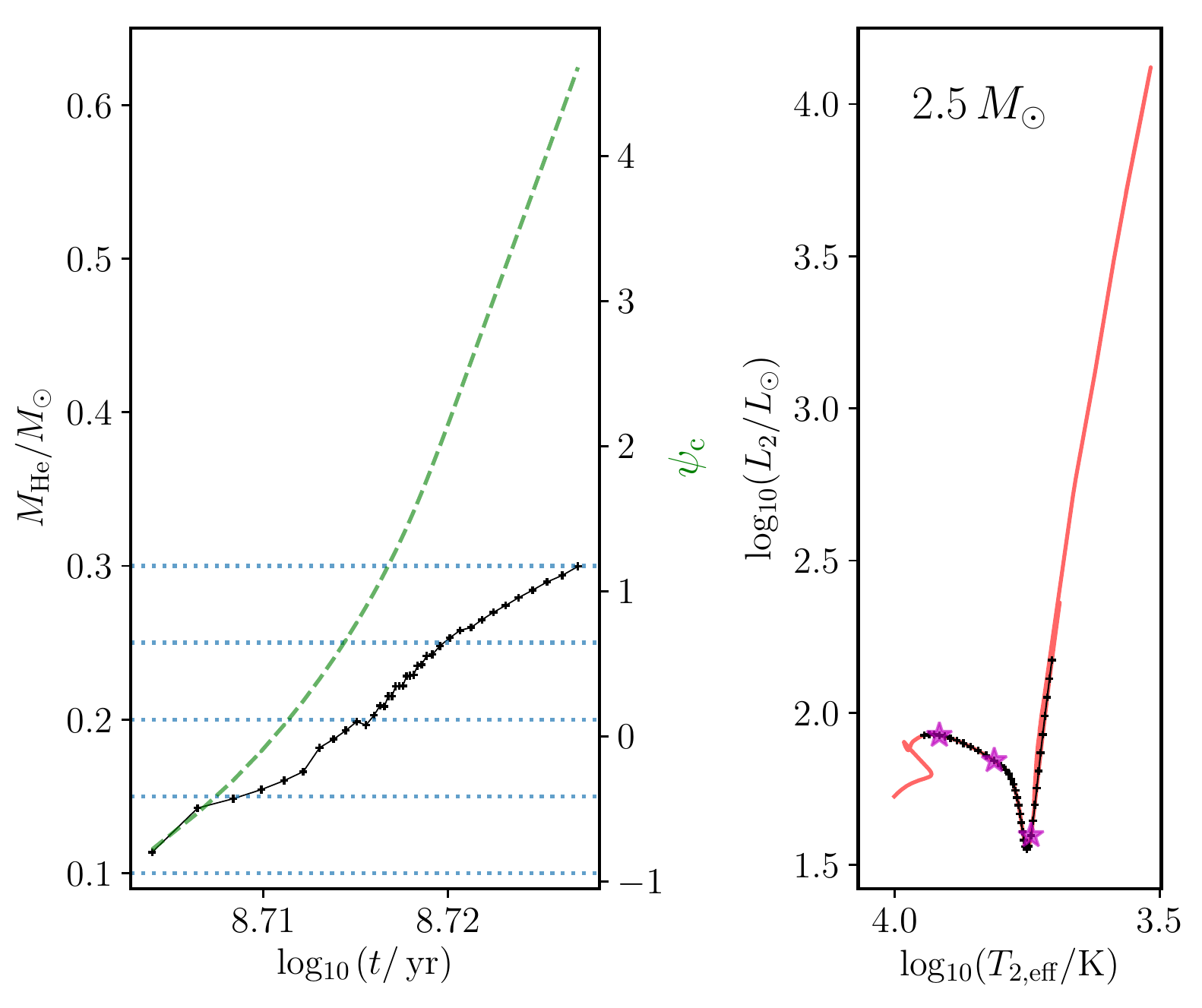}
\caption{Similar to Fig.~\ref{fig:ce_1p5}, but for a $2.5M_\odot$ secondary star.}
\label{fig:ce_2p5}
\end{figure}

\begin{figure}
\includegraphics[width=0.5\textwidth]{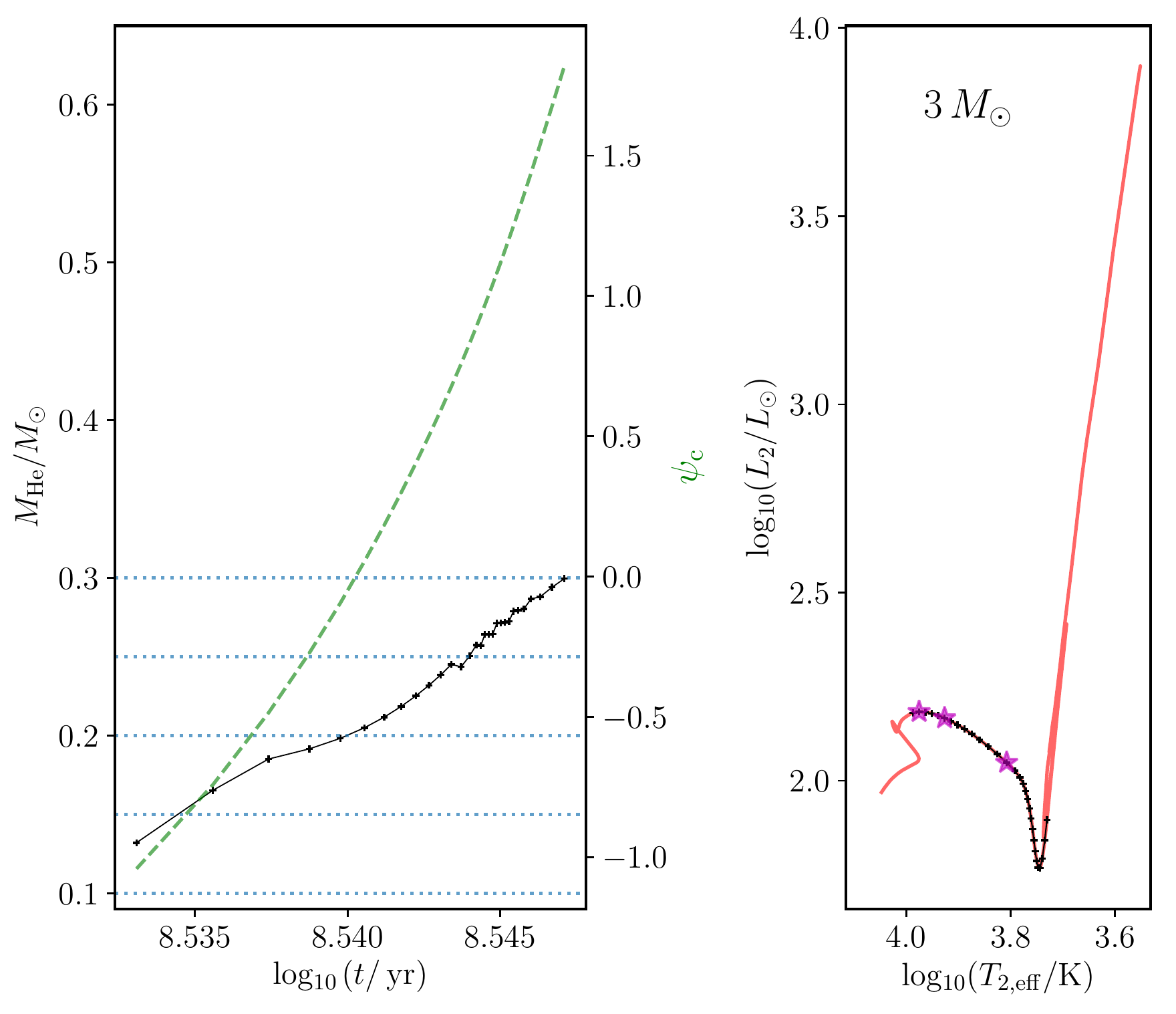}
\caption{Similar to Fig.~\ref{fig:ce_1p5}, but for a $3M_\odot$ secondary star.}
\label{fig:ce_3}
\end{figure}
The viable range of $\alpha_\mathrm{CE}$ is extremely uncertain. When $\alpha_\mathrm{CE}\leq 1$ only the orbital energy of the cores is used to eject the envelope. However this does not lead to the full ejection of the CE in many simulations \citep{2013A&ARv..21...59I}. This has led to the introduction of additional sources of energy which may contribute to the CE ejection, such as recombination energy in the ionization zones \citep{Han1995}, nuclear burning on the surface of the WD accretor \citep{Iben1993}, dust-driven winds \citep{Glanz2018} etc. In addition, the relative importance of these mechanisms as additional energy sources is still not well known. For instance, \cite{Igoshev2020} state that recombination can act sufficiently effectively in low-mass stars ($M_2\lesssim~3 M_\odot$) but not in more massive stars. Nuclear burning on the WD accretor can provide a significant additional energy source $E_\mathrm{nuc}$ which can be used to eject the CE. However, tapping this energy critically depends on the time $t_\mathrm{CE}$ for which the CE lasts such that $E_\mathrm{nuc}\propto t_\mathrm{CE}$. Constraints on $t_\mathrm{CE}$ are still poor, with $100\,\mathrm{yr}\lesssim t_\mathrm{CE}\lesssim 10^4\,\mathrm{yr}$ \citep{Igoshev2020}. \textcolor{black}{We do not present any detailed analyses of these additional sources of energy, but what is expected is that they, in conjunction with $\Delta E_\mathrm{orb} = \lvert E_\mathrm{orb,f} - E_\mathrm{orb,i}\rvert$, reduce the energy burden on orbital energy to eject the CE.} \textcolor{black}{It is also possible that CEE proceeds in a different way than the classical energy formalism, as prescribed by \citet[see also Section~\ref{subs:candidates}]{Hirai2022} which, in the classical framework corresponds to using larger values of $\alpha_\mathrm{CE}$.} So we mimic their effect by increasing the maximum allowed $\alpha_\mathrm{CE}$ beyond unity.  

We model the CEE outcome for four progenitor stars, $M_2/M_\odot \in \{1.5,\,2,\,2.5,\,3\}$ that undergo CE evolution with a WD primary of mass $M_1 = 0.9M_\odot$\footnote{This choice of the WD primary was motivated by the observed accretor masses of Gaia14aae and ZTFJ1637+49 (Table~\ref{tab:table}).} and emerge from the CE with a H-exhausted core of mass $M_\mathrm{He}$ such that $0.1\leq M_\mathrm{He}/M_\odot \leq 0.3$. \textcolor{black}{The mass ratio $q$ for these systems is greater than the critical mass ratio for dynamical time-scale mass transfer for most evolutionary stages of the donor star (see figures~6 and 7 of \citealt{Ge2020}). For the system with $M_2 = 1.5M_\odot$, our detailed models show that the binary undergos delayed dynamically unstable mass transfer when RLOF commences in the subgiant phase of the donor. The evolution of $M_\mathrm{He}$ for the different progenitors is shown in Figs~\ref{fig:ce_1p5}, \ref{fig:ce_2}, \ref{fig:ce_2p5} and \ref{fig:ce_3}. We also plot the evolution of the central degeneracy parameter $\psi_\mathrm{c}$ with time because it critically governs the trajectory of the binary ($M_\mathrm{He},\,M_1$) after it emerges from CEE (Section~\ref{sec:models}). Our motivation for choosing this set of progenitors is an interplay between $\psi_\mathrm{c}$ of the eventual He-star and the feasibility of CE. On the one hand the more massive the star, the lower the $\psi_\mathrm{c}$ of the emerging He-star. Less massive stars lead to highly degenerate donors which behave as He WDs rather than He-stars which, in our analysis, need to be semi-degenerate (Section~\ref{sec:models}). On the other hand the more massive the star the less feasible it is for the $(M_2,\,M_1)$ system to emerge from CE with a small ($0.1\leq M_\mathrm{He}/M_\odot \leq 0.3$) He-rich donor. This is because more massive stars form a He-rich core on their subgiant branch, where their radius $R_2$ is smaller than on the RGB. A smaller $R_2$ increases $E_\mathrm{bind}$ and makes it more energetically burdensome for the system to eject the CE. This is evident in equations~(\ref{eq:cee_ebind1}) and (\ref{eq:cee_ebind2}).} 
As discussed before, we use $a_\mathrm{f}$ as an independent variable such that after the CE ejection, the $(M_\mathrm{He},\,M_1)$ pair emerges with a  final orbital period $P_\mathrm{orb,f}/\mathrm{min} \in [10,\,100]$. We can then find $a_\mathrm{f}$ using Kepler's third law

\begin{center}
\begin{equation}
\label{eq:cee_k3}
a_\mathrm{f}^3 = P_\mathrm{orb,f}^2 G \frac{M_1 + M_\mathrm{He}}{4\pi^2}.
\end{equation}
\end{center}
We evolve the star with mass $M_2$ using \textsc{STARS} and track $R_2$, $M_\mathrm{He}$ and $R_\mathrm{He}$ with time. Here $R_\mathrm{He}$ is defined as the radius of the H-exhausted core. For each $M_2$, $M_\mathrm{He}(t)$, $R_\mathrm{He}(t)$ and $R_2(t)$ we calculate $\alpha_\mathrm{CE_1}$ and $\alpha_\mathrm{CE_2}$ using equations (\ref{eq:cee_alpha1}), (\ref{eq:cee_alpha2}) and (\ref{eq:cee_rl}) and check whether $\alpha_\mathrm{CE_1}$ falls into one of three ranges $0<\alpha_\mathrm{CE_{1}}\leq1$, $1<\alpha_\mathrm{CE_{1}}\leq3$, and $3<\alpha_\mathrm{CE_{1}}\leq6$ and $\alpha_\mathrm{CE_2}$ falls into one of three ranges $0<\alpha_\mathrm{CE_{2}}\leq1$, $1<\alpha_\mathrm{CE_{2}}\leq3$, and $3<\alpha_\mathrm{CE_{2}}\leq4$ (see Section \ref{subs:candidates} for the justification of the upper limits on $\alpha_\mathrm{CE_{1\;and\;2}}$). \textcolor{black}{Because the binary must emerge detached from the CEE, we also ensure that $R_\mathrm{L,He}\geq R_\mathrm{He}$, where $R_\mathrm{L,He}$ is the Roche lobe radius of the binary with stars of mass $M_\mathrm{He}$ and $M_1$. This condition does not always ensure that the binary emerges detached because we have seen that semi-degenerate He stars expand when their envelope is removed and thus their radius after CEE may be greater than $R_\mathrm{He}$. However it is a fairly good estimate for a lower cut-off for the final orbital period of many systems\footnote{In any case, any model that is wrongly assumed to be viable at this point fails to evolve with STARS.} .} If a model satisfies these conditions, we say that the system can emerge from the CEE with a tuple $(M_\mathrm{He},\,P_\mathrm{orb,f},\,M_1)$. Otherwise the system merges.

\subsection{Viable candidates after CEE}
\label{subs:candidates}

\begin{figure}
\includegraphics[width=0.5\textwidth]{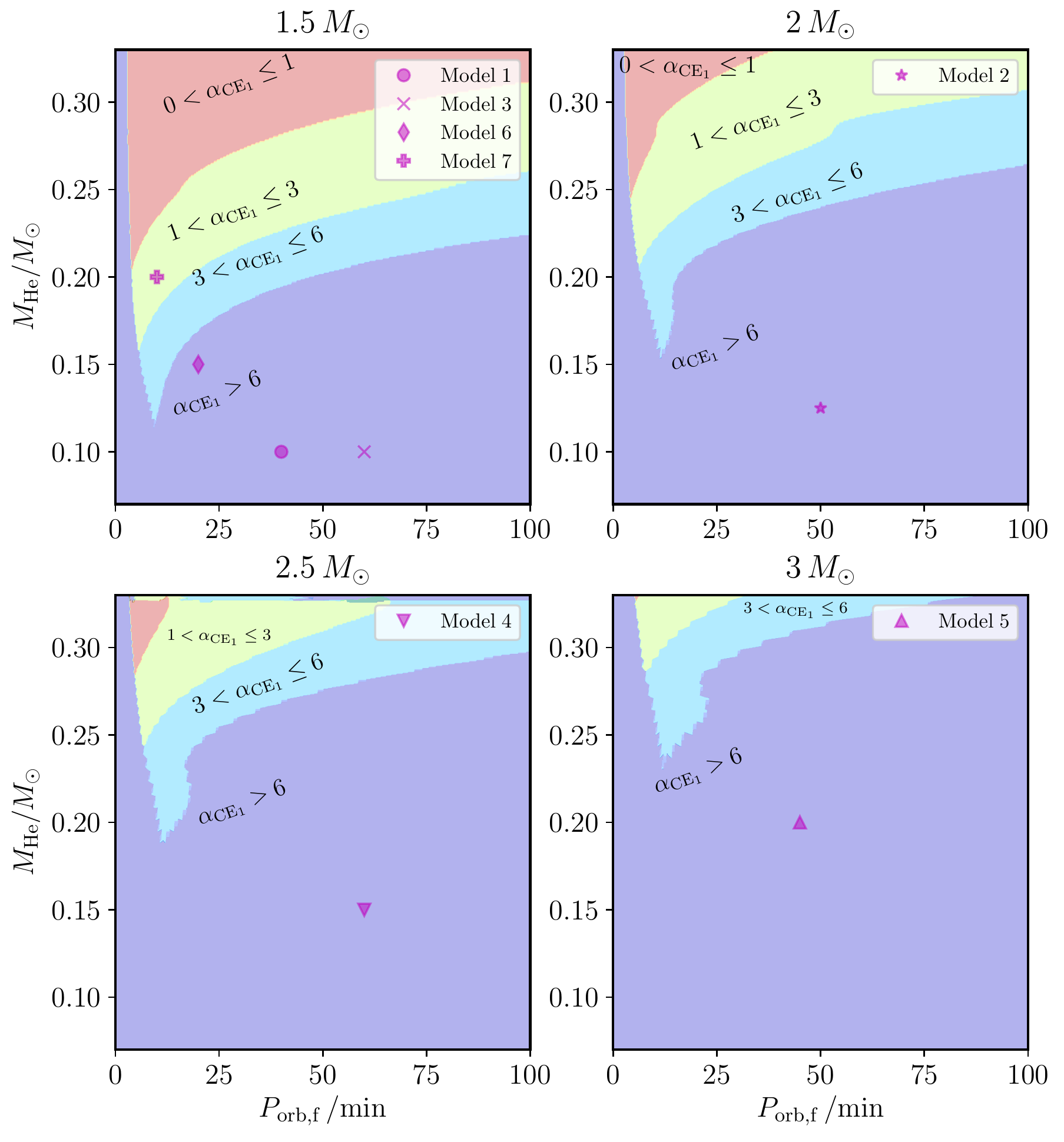}
\caption{Contour plots showing viable pairs ($P_\mathrm{orb,f},\,M_\mathrm{He}$) after CEE with a $0.9M_\odot$ WD accretor. The region in red denotes $0<\alpha_\mathrm{CE_1}\leq1$, the region in green denotes $1<\alpha_\mathrm{CE_1}\leq3$ and the region in cyan denotes $3<\alpha_\mathrm{CE_1}\leq6$. The dark blue region denotes candidates for which $\alpha_\mathrm{CE_1}>6$ and so are not feasible candidates in our analysis. Here $\alpha_\mathrm{CE_1}$ is for $E_\mathrm{bind}$ as used by \protect\citet[equation~\ref{eq:cee_ebind1}]{Tout1997}. \textcolor{black}{The initial conditions of the models of the AM CVn progenitors from Fig.~\ref{fig:prm} are shown as magenta symbols.}}
\label{fig:alpha1}
\end{figure}

\begin{figure}
\includegraphics[width=0.5\textwidth]{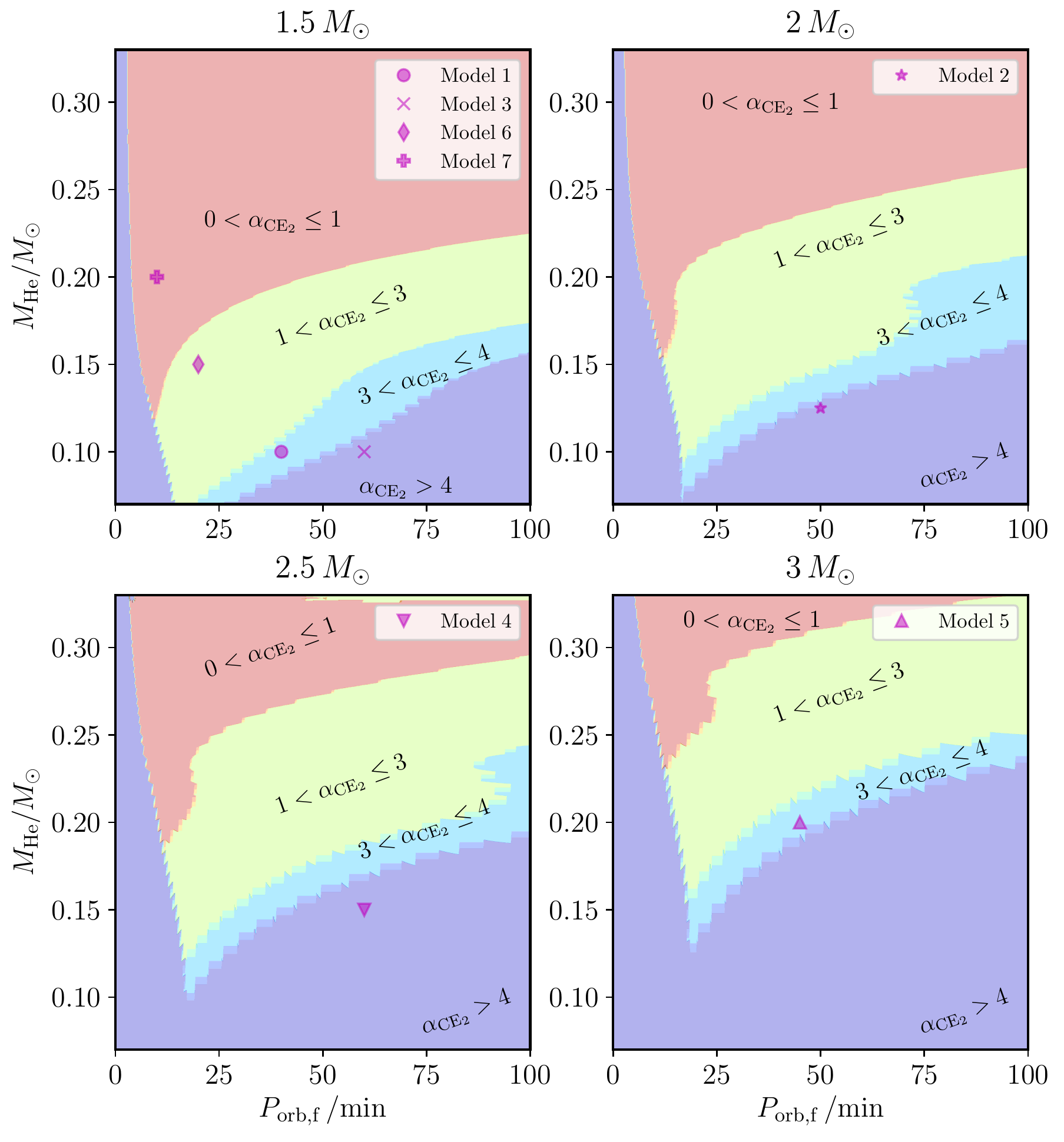}
\caption{Contour plots showing viable pairs ($P_\mathrm{orb,f},\,M_\mathrm{He}$) after CEE with a $0.9M_\odot$ WD accretor. The region in red denotes $0<\alpha_\mathrm{CE_2}\leq1$, the region in green denotes $1<\alpha_\mathrm{CE_2}\leq3$ and the region in cyan denotes $3<\alpha_\mathrm{CE_2}\leq4$. The dark blue region denotes candidates for which $\alpha_\mathrm{CE_2}>4$ and so are not feasible candidates in our analysis. Here $\alpha_\mathrm{CE_2}$ is for $E_\mathrm{bind}$ as used by \protect\citet[equation~\ref{eq:cee_ebind2}]{Iben1993}. \textcolor{black}{The initial conditions of the models of the AM CVn progenitors from Fig.~\ref{fig:prm} are shown as magenta symbols.}}
\label{fig:alpha2}
\end{figure}

Our results of viable post-CEE candidates for the different progenitors are shown in Figs~\ref{fig:alpha1} and \ref{fig:alpha2}, for which the mass of the WD accretor is kept fixed at $0.9M_\odot$. \textcolor{black}{In each plot, the region in red denotes $0<\alpha_\mathrm{CE_{1\,or\,2}}\leq1$, and the region in green $1<\alpha_\mathrm{CE_{1\,or\,2}}\leq3$. The region in cyan in Fig.~\ref{fig:alpha1} denotes $3<\alpha_\mathrm{CE_1}\leq6$, whereas in Fig.~\ref{fig:alpha2} it denotes $3<\alpha_\mathrm{CE_1}\leq4$. The dark blue region in the figures denote candidates for which $\alpha_\mathrm{CE_1}>6$ and $\alpha_\mathrm{CE_2}>4$, implying that these candidates are unfeasible energetically, and that these systems merge.} 
It can be seen that, in general, the larger the maximum allowed $\alpha_\mathrm{CE}$, the more systems emerge from CEE successfully. This can be explained because a larger allowed $\alpha_\mathrm{CE}$ mimics a larger reservoir of additional energy sources which can be used to overcome the envelope's binding energy. It can also be seen that, in general, more systems emerge from CEE successfully with $\alpha_\mathrm{CE_2}$ than with $\alpha_\mathrm{CE_1}$. This can be attributed to the fact that equation (\ref{eq:cee_ebind2}) yields a lower binding energy of the envelope than equation (\ref{eq:cee_ebind1}) by a factor 
\begin{center}
\begin{equation}
\label{eq:cee_alphar}
\frac{\alpha_\mathrm{CE_1}}{\alpha_\mathrm{CE_2}} = \frac{4}{0.49}\frac{q}{1 + q}\frac{0.6q^{2/3} + \mathrm{ln}(1 + q^{1/3})}{q^{2/3}}
\end{equation}
\end{center}
which, for $M_2/M_\odot \in \{1.5,\,2,\,2.5,\,3\}$ and $M_1 = 0.9M_\odot$ is about $\{5.90,\,6.14,\,6.27,\,6.34\}$. \textcolor{black}{The red region in both figures corresponds to a maximum allowed $\alpha_\mathrm{CE} =1$, which is equivalent to only the orbital energy of the core unbinding the envelope.} As can be seen, fewer systems emerge successfully if only this energy reservoir is used, with even fewer for more massive stars. The \textcolor{black}{green and cyan regions} have progressively larger maxima allowed for $\alpha_\mathrm{CE}$s. The choice of these $\alpha_\mathrm{CE}$s is rather arbitrary because there is uncertainty in the efficiency with which additional energy sources contribute to the ejection of the CE. In general $\alpha_\mathrm{CE_1}/\alpha_\mathrm{CE_2}\approx6$ for our progenitors so $\alpha_\mathrm{CE_1}=6$ mimics a maximum $\alpha_\mathrm{CE_2}=1$, leading to almost the same tuples successfully emerging for these two cases. \textcolor{black}{We note importantly that equation~(\ref{eq:cee_ebind1}) uses a fixed value of $\lambda=0.5$ throughout the evolution of the donor star. This is not true, as pointed out by \cite{1994MNRAS.270..121H,2000A&A...360.1043D} etc. \cite{2010ApJ...716..114X} show the evolution of $\lambda$ with stellar radius (see their fig.~1), where it can be seen that for most of the donor progenitors with masses and radii of our interest, $\lambda \gtrsim 1$. Factoring in this will reduce $\alpha_\mathrm{CE_1}$ which, in our formalism, is equivalent to setting the maximum $\alpha_\mathrm{CE_1}>1$. For the analysis in this work, we choose $\alpha_\mathrm{CE_{1}}=6$ as the largest allowed CE efficiency parameter for the formalism of \cite{Tout1997}. } 

\textcolor{black}{We now touch upon a different approach to determine the post-CE system properties, described in detail in \cite{Hirai2022} for red supergiants. They argue that red supergiants have a sizeable radiative zone between the convective envelope and the He core, and after the commencement of CE, the spiral-in through the convective layer can be treated as a dynamical time-scale process. However, once the radiative layers are reached, the mass transfer occurs on a thermal time-scale and the assumption of energy conservation breaks down as nuclear burning on the thermal time-scale becomes comparable to the binding energy. For such systems the CEE should be modelled with what they call a two-step formalism. However, low-mass giants have a negligible radiative layer and the whole CEE can be simply modelled as a dynamical time-scale phenomenon. Thus, they show that the classical energy formalism for CEE is inadequate for modelling the CE phase in red supergiants. We now explain how their result is relevant in our work. We showed in Section~\ref{subs:method} that in order to emerge as a semi-degenerate He-star with $0.1\leq M_\mathrm{He}/M_\odot \leq 0.3$, the progenitor donor must commence unstable RLOF between its Hertzprung-gap (subgiant) phase and the red giant phase, with more contribution from subgiant donors for larger $M_2$. These subgiants have an internal structure quite similar to red supergiants, with a He core, a sizeable radiative layer, and a convective envelope. Therefore, the treatment of CEE for these systems should be made under the two-step formalism. Working out the post-CE separations for our progenitor systems under this formalism is beyond the scope of this work, and we simply generalise the result of \cite{Hirai2022} which states that the two-step formalism yields larger post-CE orbital separations than the classical formalism. In the classical energy formalism, this is equivalent to setting $\alpha_\mathrm{CE}\gtrsim10$ (see their section~3). \cite{Hirai2017} deduce an even larger $\alpha_\mathrm{CE}\gtrsim20$ for the Type Ib supernova iPTF13bvn. In order to assess the difference between the two different formalisms of the treatment of CEE, we set an ad hoc maximum $\alpha_\mathrm{CE_1}=24$, which yields $\alpha_\mathrm{CE_2}\approx4$ for the binding energy description of \citet[according to equation~\ref{eq:cee_alphar}]{Iben1993}. So with much larger allowed value of $\alpha_\mathrm{CE}$, Fig.~\ref{fig:alpha2} describes the viable post-CE candidates under the two-step treatment of CEE, while Fig.~\ref{fig:alpha1} describes the viable post-CE candidates under the classical treatment of CEE with additional sources of energy. We can see that more models of AM CVn progenitors of our interest (from Fig.~\ref{fig:prm}) are viable under the two-step formalism compared to the solely classical energy formalism. We urge for a more rigorous study of the evolution of these pre-CE systems under the two-step formalism. }

\section{Detailed models}
\label{sec:models}
In this section, we use \textsc{STARS} to make detailed models of AM CVn systems formed through the He-star channel, after emerging from the CE. Hereinafter, owing to uncertainties in modelling the CE outcome in detail, we do not comment on the relative likelihood of any particular system emerging successfully from a CE. From here on we label the He-star, the mass-losing secondary donor, by He and denote its mass and radius by $M_\mathrm{He}$ and $R_\mathrm{He}$. We assume that, upon commencement of RLOF, the system undergoes fully non-conservative mass transfer from the donor to the WD accretor such that all the mass accreted by the primary is expelled in the form of nova outbursts which carry away the specific angular momentum of the WD accretor. This is a reasonable assumption because the mass-transfer rate during the entire evolution is much lower than $10^{-6}\,M_\odot\,\mathrm{yr}^{-1}$, which is approximately the stable burning regime for He-dominated matter \citep{1982ApJ...253..798N}.

We evolve our systems with $\mathrm{AML_{GR}}$ (see \citealt{1981ApJ...248L..27P} and the references therein) given by 
\begin{center}
\begin{equation}
\label{eq:gr}
\frac{\Dot{J}_\mathrm{GR}}{J} = -32G^3{{M_1M_\mathrm{He}}}\frac{M_1+M_\mathrm{He}}{5a^4c^5},
\end{equation}
\end{center}
where $a$ is the orbital separation of the binary and $c$ is the speed of light. This has been the primary mechanism for AML used by many groups (e.g. \citealt{Yungelson2008} on the He-star Channel) for short orbital periods. However, we evolve our trajectories with an additional source of AML, described in detail by ST, wherein the interplay between two $\alpha-\Omega$ dynamos in a low-mass star gives rise to stellar winds from the donor and leads to AML \citep{1987MNRAS.226...57M}. We call this the Double Dynamo (DD) mechanism of angular momentum loss ($\mathrm{AML_{DD}}$). \textcolor{black}{An $\alpha-\Omega$ dynamo is a mechanism with which a differentially rotating body can sustain long-term magnetic fields. The first dynamo operates in the outer convective region of the donor, wherein shear in the envelope converts poloidal field to toroidal field (the $\Omega$ term in the $\alpha-\Omega$ dynamo model) and convection converts toroidal field to poloidal field (the $\alpha$ term in the $\alpha-\Omega$ dynamo model). This has been discussed in detail by \cite{1992MNRAS.256..269T}. The other dynamo operates in the boundary layer between the corotating convective envelope and the slowly rotating radiative core. Here the $\alpha-\Omega$ dynamo is driven by the strong differential rotation between the core and the envelope \citep{Zangrilli1997}. } The mathematical formulation of $\mathrm{AML_{DD}}$ is given by equations (3) to (25) of ST and we urge the reader to refer to it for a detailed explanation of the relevant terms\footnote{We point out that $M_2$ in ST is the mass of the donor star. Here it is $M_\mathrm{He}$.}. For clarity the primary expression for $\mathrm{AML_{DD}}$ is
\begin{center}
\begin{equation}
\label{eq:jdotbyjdd}
\frac{\Dot{J}_\mathrm{DD}}{J} \propto \Dot{M}_\mathrm{w}\frac{M_1+M_\mathrm{He}}{M_1M_\mathrm{He}}\left(\frac{R_\mathrm{A}}{a}\right)^2,
\end{equation}
\end{center}
where $\Dot{M}_\mathrm{w}$ is the donor's mass-loss rate in its wind and the Alfvén radius of the donor
\begin{center}
\begin{equation}
\label{eq:ra}
R_\mathrm{A} = R_\mathrm{He} \left(\frac{B_\mathrm{p}^2R_\mathrm{He}^2}{\Dot{M}_\mathrm{w}v_\mathrm{w}}\right)^{1/4},
\end{equation}
\end{center}
where $v_\mathrm{w}$ is the escape velocity of the donor and $B_\mathrm{p}$ is the poloidal component of the magnetic field of the donor. In our DD model, $\Dot{M}_\mathrm{w}$ and $B_\mathrm{p}$ are given by
\begin{center}
\begin{equation}
\label{eq:mltot}
\Dot{M}_\mathrm{w} = \Dot{M}_\mathrm{w,conv} + \Dot{M}_\mathrm{w,bl}
\end{equation}
\end{center} and 
\begin{center}
\begin{equation}
\label{eq:bptot}
B_\mathrm{p} = B_\mathrm{p,conv} + B_\mathrm{p,bl},
\end{equation}
\end{center}
where the subscripts `$\mathrm{conv}$' and `$\mathrm{bl}$' denote the contributions by the convective and the boundary layer dynamos respectively. ST use three free parameters $(\alpha,\,\beta,\,\gamma) = (4.6, 0.08, 3.2)$ to model the complete evolution of CVs from the beginning of RLOF to beyond the period minimum (see \citealt{2003cvs..book.....W} for a detailed review of CVs). \textcolor{black}{These free parameters govern the enhancement of the convective dynamo for CVs below the period gap, the efficiency of the boundary layer dynamo, and the efficiency of orbital angular momentum loss respectively. The values of $\beta$ and $\gamma$ were chosen so that the period gap is well reproduced in canonical CV tracks, and $\alpha$ was chosen to reproduce the correct period minimum spike in CV distribution.} We choose the same parameters here. We show in Section~\ref{sec:dd} that none of our selected trajectories evolving with only $\mathrm{AML_{GR}}$ match with either Gaia14aae or ZTFJ1637+49. We also argue why an extra AML mechanism ought to be at play in these systems.

\subsection{Analysis of observed parameters}
\label{subs:param}

\begin{figure*}
\centering
\includegraphics[width=1\textwidth]{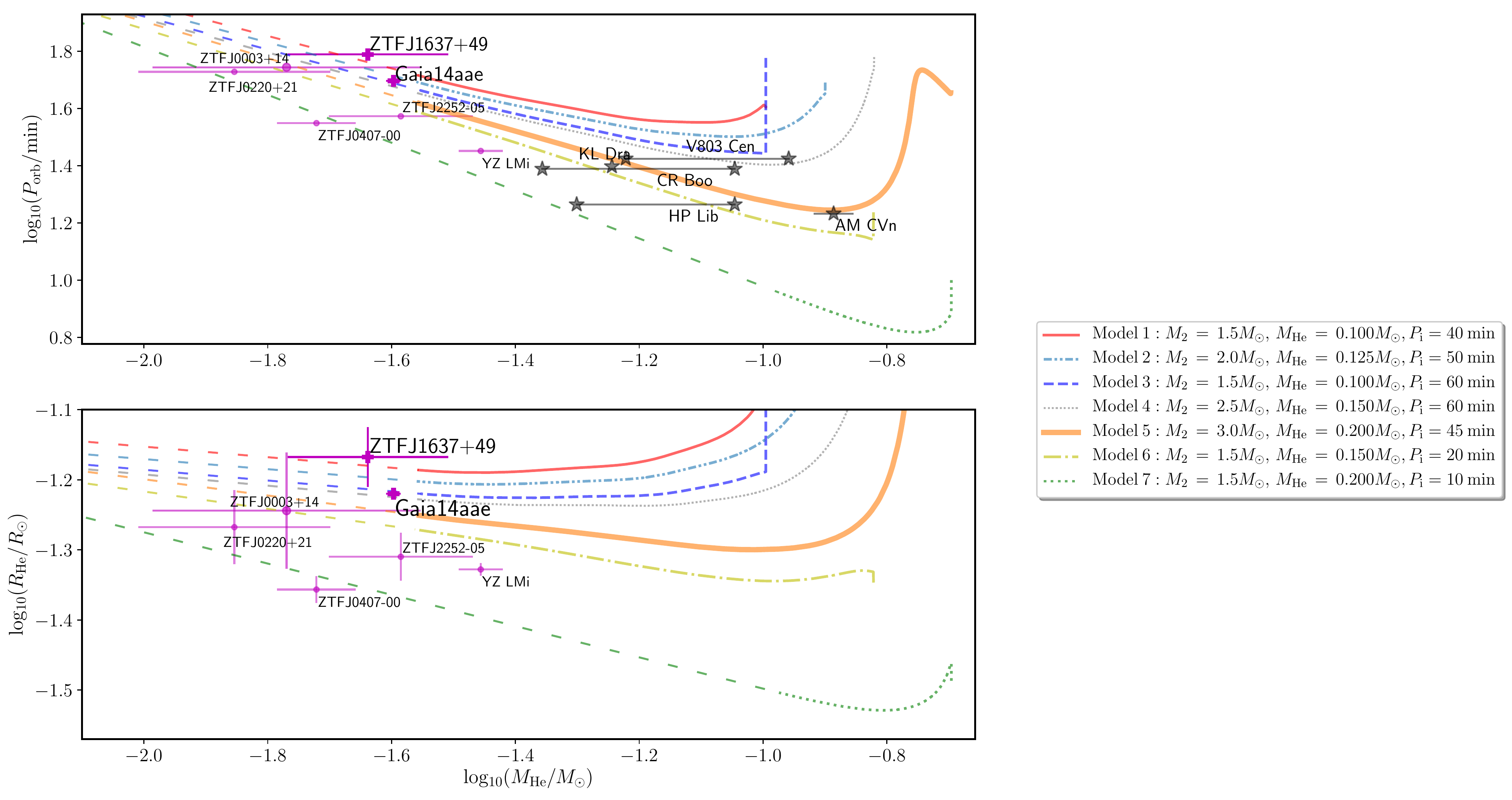}
\caption{The evolutionary tracks of our He-stars in the $(\mathrm{log}\,M_\mathrm{He},\,\mathrm{log}\,P_\mathrm{orb})$ and $(\mathrm{log}\,M_\mathrm{He},\,\mathrm{log}\,R_\mathrm{He})$ planes, with a $0.9M_\odot$ WD primary. The dashed section in each trajectory is a power law fit of the form $P_\mathrm{orb} \propto (M_\mathrm{He}/M_{\mathrm{He},P_\mathrm{orb,mini}})^{\delta_1}$ or $R_\mathrm{He} \propto (M_\mathrm{He}/M_{\mathrm{He},R_\mathrm{mini}})^{\delta_2}$. \textcolor{black}{The trajectories evolve from right to left.} The circles and crosses in magenta are AM CVn systems described by \protect\cite{2022MNRAS.512.5440V}, \protect\cite{Green2018}, \protect\cite{Copperwheat2010}. \textcolor{black}{The crosses, Gaia14aae and ZTFJ1637+49, are the main focus of this work.} The stars in black are AM CVn systems with a He-star formation channel reported by \protect\cite{Solheim2010}.}
\label{fig:prm}
\end{figure*}

\begin{figure}
\includegraphics[width=0.5\textwidth]{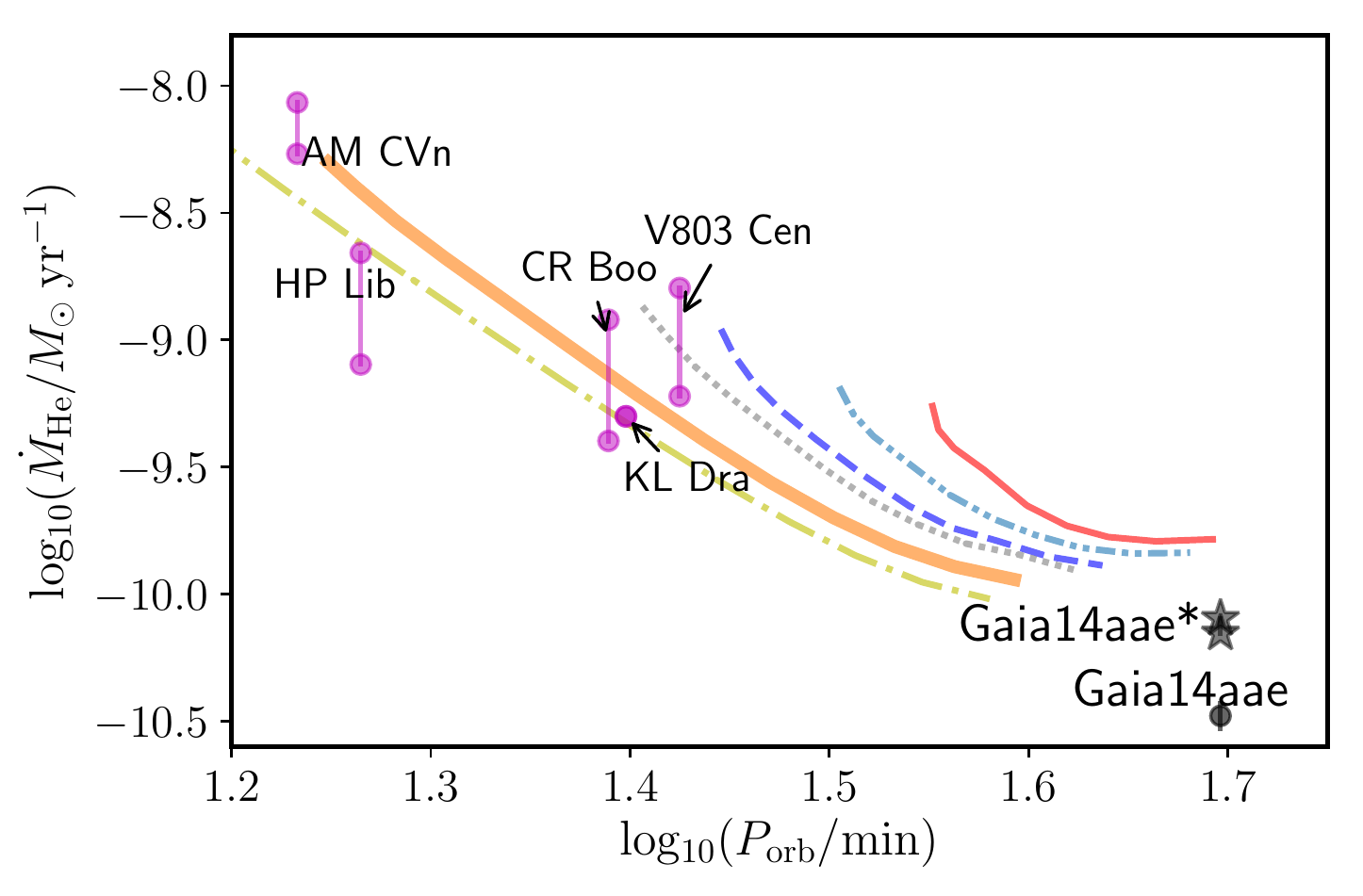}
\caption{The same models and observed systems as in Fig.~\ref{fig:prm} but in the
$(\mathrm{log}\,P_\mathrm{orb},\mathrm{log}\,\Dot{M}_\mathrm{He})$ plane. Each trajectory is modelled after the system attains its period minimum \textcolor{black}{and evolves from left to right.}. The black point labelled `Gaia14aae' corresponds to a mass-transfer rate of $\mathrm{log_{10}}(\Dot{M}_\mathrm{He}/M_\odot\,\mathrm{yr^{-1}})=-10.48 \pm 0.06$ \protect\citep{2018A&A...620A.141R} and black stars labelled `Gaia14aae*' correspond to a mass-transfer rate of $\Dot{M}_\mathrm{He}=7-8\times10^{-11} M_\odot\,\mathrm{yr^{-1}}$ \protect\citep{Campbell2015}.}
\label{fig:mdotp}
\end{figure}

Using $\mathrm{AML_{DD}}+\mathrm{AML_{GR}}$, we compare our detailed trajectories to the mass, radius and mass-transfer rate of the donor and the orbital period of all eclipsing AM CVn systems reported by \citet[and the references therein]{2022MNRAS.512.5440V}, including Gaia14aae and ZTFJ1637+49, which are the main focus of this work. We also consider well-known AM CVn systems with a proposed formation through the He-star channel reported by \cite{Solheim2010}. The initial orbital periods and masses $P_\mathrm{i}$, $M_2$ and $M_\mathrm{He}$ are chosen to demonstrate the effect of $\psi_\mathrm{c}$ and $M_2$ on our results. The central degeneracy $\psi_\mathrm{c}$ governs the evolution of $M_\mathrm{He}$, $R_\mathrm{He}$,  $\Dot{M}_\mathrm{He}$ and  $P_\mathrm{orb}$, while $M_2$ governs the abundances of the donor which, in turn, govern the abundances of the accretion disc. This is shown in Figs~\ref{fig:prm} and \ref{fig:mdotp}, \textcolor{black}{where the models are named in an increasing order of their central degeneracy at $P_\mathrm{orb,mini}$.} While, with STARS, we are unable to model the complete evolution of the donor, below about $0.03 M_\odot$\footnote{This is because in STARS, mass loss from degenerate layers of a star is unstable. As a result some trajectories involving highly degenerate donors such as the green model in Fig.~\ref{fig:prm} cannot be evolved below $0.03M_\odot$. We shall overcome this limitation in the future but do not believe that it detracts from our conclusions here.}, we observe that in the $(M_\mathrm{He},P_\mathrm{orb})$ and $(M_\mathrm{He},R_\mathrm{He})$ planes, the trajectories follow a power law relation after their minima in $P_\mathrm{orb}$ and $R_\mathrm{He}$. So we can extrapolate the trajectories for smaller $M_\mathrm{He}$ with simple power-law relations of the form
\begin{center}
\begin{equation}
\label{eq:models_delta1}
P_\mathrm{orb}(M_\mathrm{He}) = P_\mathrm{orb}(M_{\mathrm{He},P_\mathrm{orb,mini}}) \left(\frac{M_\mathrm{He}}{M_{\mathrm{He},P_\mathrm{orb,mini}}}\right)^{\delta_1}
\end{equation}
\end{center}
and
\begin{center}
\begin{equation}
\label{eq:models_delta2}
R_\mathrm{He}(M_\mathrm{He}) = R_\mathrm{He}(M_{\mathrm{He},R_\mathrm{He,mini}}) \left(\frac{M_\mathrm{He}}{M_{\mathrm{He},R_\mathrm{He,mini}}}\right)^{\delta_2},
\end{equation}
\end{center}
where $M_{\mathrm{He},P_\mathrm{orb,mini}}$ and $M_{\mathrm{He},R_\mathrm{He,mini}}$ are the masses of the He-star when the model attains minima in $P_\mathrm{orb}$ and $R_\mathrm{He}$ respectively. We obtain $\delta_1$ and $\delta_2$ separately for each model by fitting the last 100 models with this power law. We see that our trajectories match well with the observed properties of both ZTFJ1637+49 (\textcolor{black}{Model 1 and 2}) and Gaia14aae (\textcolor{black}{Model 3 and 4}). However, we do not get conclusive agreement of our models with observations of Gaia14aae in the $(\Dot{M}_\mathrm{He},\,P_\mathrm{orb})$ plane. \textcolor{black}{This is because we do not observe a clear linear trend between $P_\mathrm{orb}$ and $\Dot{M}_\mathrm{He}$ in the log-log plane and so cannot deduce a power-law relation between them and extrapolate the model trajectories to longer periods. The absence of this trend is because of the increasing dominance of AML owing to the convective dynamo as the donor star becomes increasingly convective (Section~\ref{sec:dd}). We also point out that different mass-transfer rates have been inferred by different groups as shown in Fig.~\ref{fig:mdotp}. `Gaia14aae' corresponds to an estimate by \cite{2018A&A...620A.141R}, while the estimate by \cite{Campbell2015} is shown as `Gaia14aae*'. We see that our models agree better with the latter and urge that more precise measurements of its mass-transfer rate be made.}

\subsection{Analysis of abundances}
\label{subs:abun}

\begin{figure}
\includegraphics[width=0.4\textwidth]{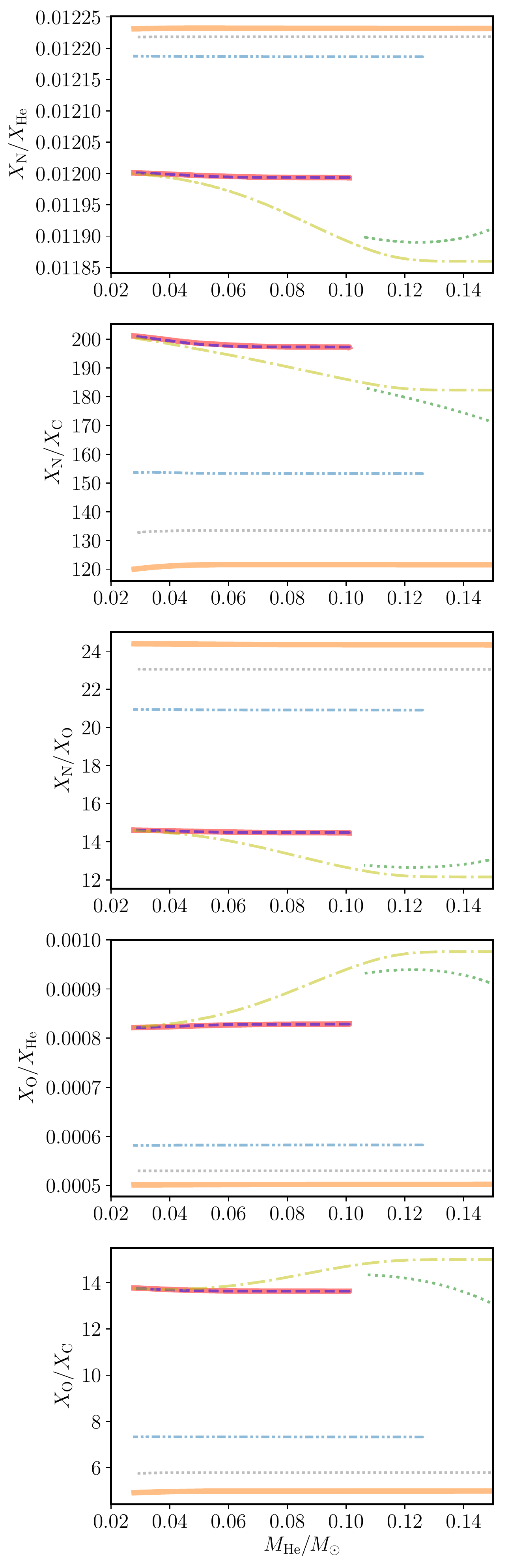}
\caption{The evolution of surface mass-fraction ratios of C, N, O and He with the He-star mass for the same set of systems as in Fig.~\ref{fig:prm}. \textcolor{black}{The trajectories evolve from right to left.} \textcolor{black}{The dark blue dashed line and the red solid line are coincident.}  }
\label{fig:CNO}
\end{figure}

We also model the evolution of the mass-fractions of C, N, O and He of the donor with mass. We look at the mass-fraction ratios described by \citet[see their figs 11 and 12]{Nelemans2010}, shown in Fig.~\ref{fig:CNO}. Our AM CVn systems have more extreme mass-fraction ratios than theirs. We find $X_\mathrm{N}/X_\mathrm{He}\gtrsim 0.12$, $120 \lesssim X_\mathrm{N}/X_\mathrm{C}\lesssim 200$, $12 \lesssim X_\mathrm{N}/X_\mathrm{O}\lesssim 24$, $5\times 10^{-4} \lesssim X_\mathrm{O}/X_\mathrm{He}\lesssim  10^{-3}$ and $5 \lesssim X_\mathrm{O}/X_\mathrm{C}\lesssim 14$,  while they report $X_\mathrm{N}/X_\mathrm{He}\lesssim 0.1$, $X_\mathrm{N}/X_\mathrm{C}\lesssim 100$, $X_\mathrm{N}/X_\mathrm{O}\lesssim 10$, $10^{-3} \lesssim X_\mathrm{O}/X_\mathrm{He}\lesssim  40$ and $0.05 \lesssim X_\mathrm{O}/X_\mathrm{C}\lesssim 5$. Our results match better with their hybrid WD donor results. All the abundance trends in our plots and the difference in the results can be attributed to the fact that our progenitor stars do not undergo any He burning and so are not enhanced in C. The donor abundances are governed just by CNO-processing and the ratios depend on the temperature at which CNO-equilibrium is reached in the donor. In turn this depends on the mass of the progenitor. More massive stars reach CNO-equilibrium with enhanced C and N, and reduced O. Our low-mass He-stars emerge from the CE phase with their abundances frozen-in at the onset of the CEE. We also note that the AM CVn systems from the catalogue of \cite{2022MNRAS.512.5440V} and most of the systems from Table 1 of \cite{Nelemans2010} have reported detection of N but no detection of C. Our results are also consistent with the non-detection of O in a number of systems because more massive progenitors suppress O relative to C and N. Now \cite{2022MNRAS.512.5440V} detect N but no O in ZTFJ1637+49. In our analysis this can be explained better when $M_2 = 2 M_\odot$. This favours our \textcolor{black}{Model 2} over \textcolor{black}{Model 1}. However, \cite{Green2019} detect the presence of both N and O in Gaia14aae. In our analysis this can be explained better by $M_2 = 1.5M_\odot$. This favours \textcolor{black}{Model 3} over \textcolor{black}{Model 4}. 

\section{Effects of additional angular momentum loss mechanisms}
\label{sec:dd}

\begin{figure*}
\centering
\includegraphics[width=1\textwidth]{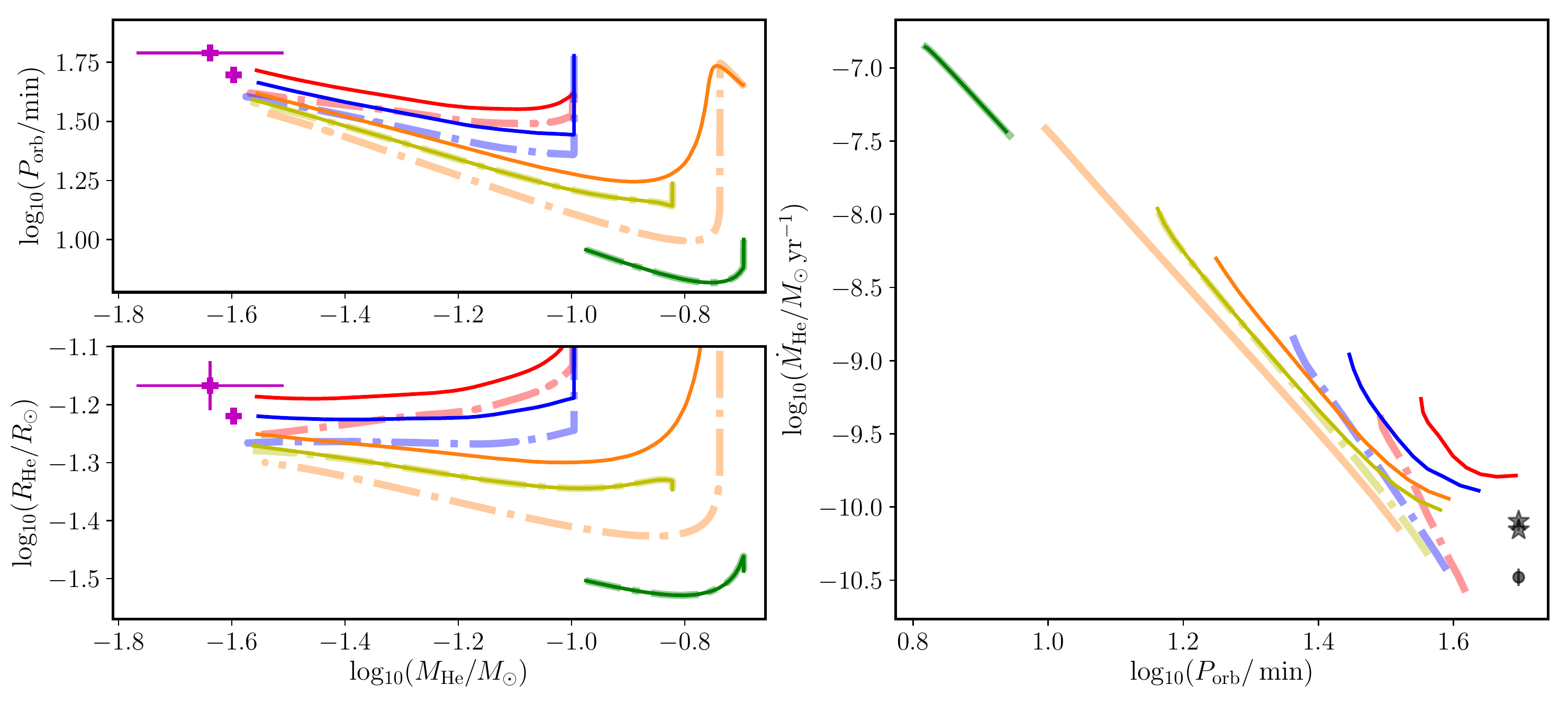}
\caption{Selected evolutionary tracks from Figs~\ref{fig:prm} and \ref{fig:mdotp} but without the power-law extrapolation. \textcolor{black}{The different line styles have been omitted for clarity.} The solid-line trajectories are binaries evolved with $\mathrm{AML_{DD}+AML_{GR}}$ and the thick dash-dotted trajectories are the same systems but evolved only with $\mathrm{AML_{GR}}$. The observed systems Gaia14aae and ZTFJ1637+49 are marked as in Figs~\ref{fig:prm} and \ref{fig:mdotp}.}
\label{fig:prmmdot_gr}
\end{figure*}

\begin{figure*}
\centering
\includegraphics[width=0.75\textwidth]{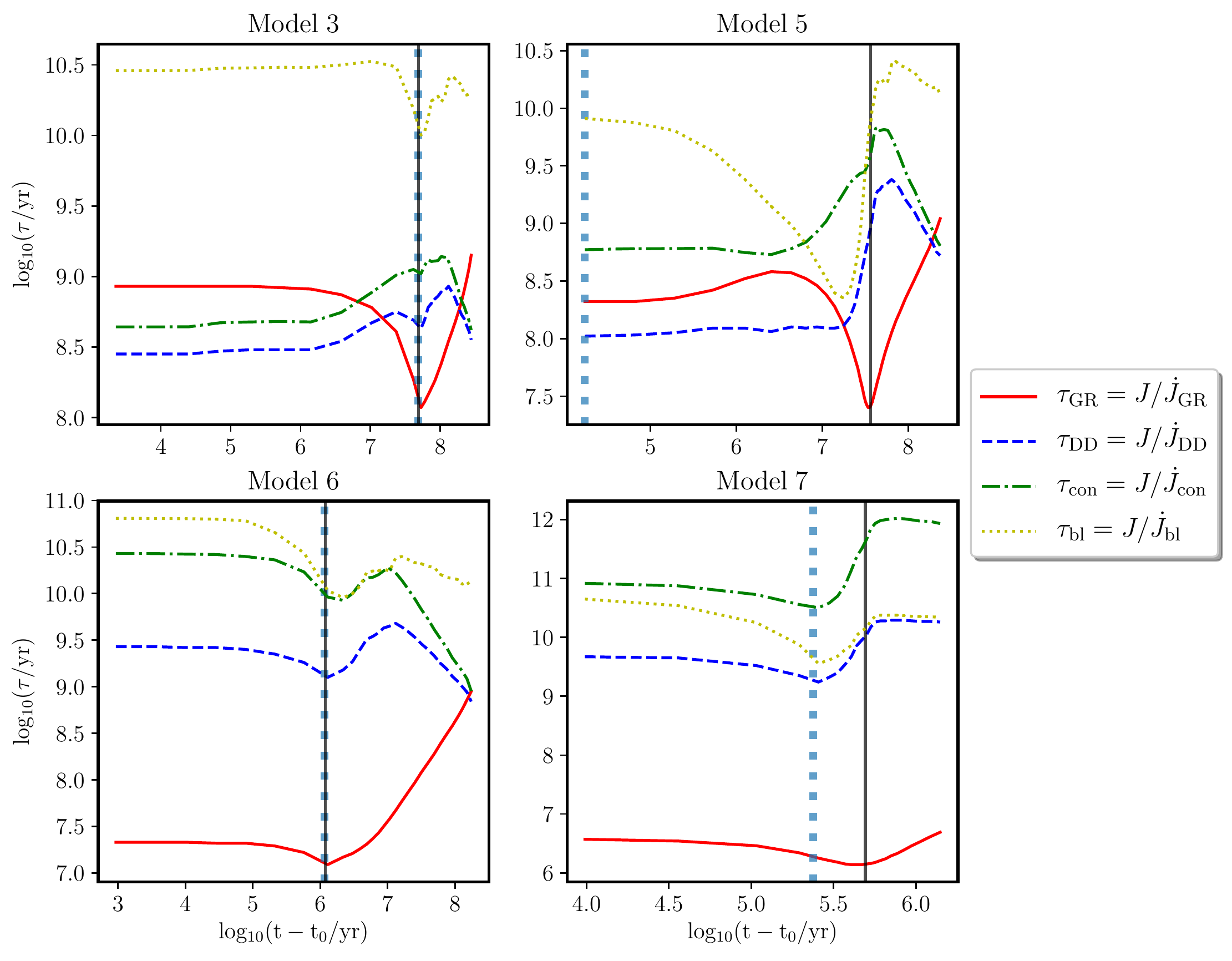}
\caption{Selected evolutionary tracks (the title of each subplot denotes the evolutionary track) from Figs~\ref{fig:prm} showing the temporal evolution of various angular momentum loss time-scales after the system emerges from CEE at $t=t_0$. The vertical light blue dash-dotted line denotes the time when RLOF commences and the vertical black solid line denotes the time when the system attains its minimum in $P_\mathrm{orb}$. Model 3 corresponds to $M_2=1.5M_\odot,\,M_\mathrm{He}=0.1M_\odot,\,P_\mathrm{i} = 60\,\mathrm{min}$, Model 5 corresponds to $M_2=3.0M_\odot,\,M_\mathrm{He}=0.2M_\odot,\,P_\mathrm{i} = 45\,\mathrm{min}$, Model 6 corresponds to $M_2=1.5M_\odot,\,M_\mathrm{He}=0.15M_\odot,\,P_\mathrm{i} = 20\,\mathrm{min}$ and Model 7 corresponds to $M_2=1.5M_\odot,\,M_\mathrm{He}=0.2M_\odot,\,P_\mathrm{i} = 10\,\mathrm{min}$.}
\label{fig:dynamo}
\end{figure*}

In the previous section we described how we employ our additional $\mathrm{AML_{DD}}$, along with $\mathrm{AML_{GR}}$, used by ST to evolve CVs. We now justify its validity. $\mathrm{AML_{DD}}$ was formulated as a physically motivated formalism to explain two prominent features of the orbital period distribution of CVs. The first is a dearth of mass-transferring systems in the region $2\,\mathrm{hr}\lesssim P_\mathrm{orb}\lesssim 3\,\mathrm{hr}$, known as the period gap. This was explained by ST using an interplay of two $\alpha-\Omega$ dynamos, namely the boundary layer dynamo and the convective dynamo, which cause angular momentum loss by magnetic braking from the donor star. The formulation of ST follows that of \cite{1983ApJ...275..713R}, wherein magnetic braking ceases to operate when the donor becomes fully convective, causing it to go out of thermal equilibrium and shrink inside its Roche lobe at $P_\mathrm{orb}\approx3\,\mathrm{hr}$. This leads to the cessation of mass-transfer, which only begins again when $\mathrm{AML_{GR}}$ has shrunk the orbit and RLOF recommences at $P_\mathrm{orb}\approx2\,\mathrm{hr}$, leading to a period gap in this $P_\mathrm{orb}$ range. ST model this because the boundary layer dynamo naturally ceases when the donor loses its boundary layer and becomes fully convective. The second feature is a peak in the number of CVs at $P_\mathrm{orb}\approx 80\,\mathrm{min}$, known as the period minimum ($P_\mathrm{orb,mini}$) spike, which arises from an interplay between the donor's thermal time-scale, mass-loss time-scale and degeneracy (\citealt{1981ApJ...248L..27P}, \citealt{1983ApJ...275..713R}). However, $P_\mathrm{orb,mini}\approx65\,\mathrm{min}$ was theoretically derived with $\mathrm{AML_{GR}}$, while observations indicated $P_\mathrm{orb,mini}\approx80\,\mathrm{min}$. This discrepancy was addressed by ST with an extra AML below the period gap owing to the convective dynamo. Using a free parameter as an enhancement factor in the convective dynamo below the period gap, ST were able to obtain good agreement with observations (see their figs 10 and 13).

The most important way in which the donors of the He-star channel differ from those of CVs is that they are H-exhausted, or in our case He-dominated, in contrast to those in CVs which begin mass transfer as MS stars with solar H abundance. The effect of evolved donors on the period gap has been studied before. \cite{1988A&A...191...57P} showed that systems with more H-exhausted (evolved) donors show a smaller period gap. This can be explained by the fact that, on the one hand, the more evolved the donor the smaller it needs to be in order to become fully convective and, on the other hand, the smaller the donor the weaker the system's angular momentum loss by magnetic braking $\mathrm{AML_{MB}}$. At some stage the time-scale of $\mathrm{AML_{MB}}$ becomes comparable to that of $\mathrm{AML_{GR}}$ for these systems and the system is only moderately out of thermal equilibrium when $\mathrm{AML_{MB}}$ ceases, leading to a smaller period gap. This also explains why highly evolved systems do not yield a period gap at all. These systems never go out of thermal equilibrium and the transition from $\mathrm{AML_{MB}}$ to $\mathrm{AML_{GR}}$ dominated evolution is smooth (see also \citealt{Podsiadlowski2002}). We argue that the physics governing $\mathrm{AML_{MB}}$ for canonical CVs should, for the most part, still be at play for evolved systems, including those with He-star donors. The way in which this extra AML affects our model trajectories is profound, as shown in Fig.~\ref{fig:prmmdot_gr}, where the thick dash-dotted lines evolved with $\mathrm{AML_{GR}}$ are the counterparts to the solid lines evolved with $\mathrm{AML_{GR}}\,+\,\mathrm{AML_{DD}}$. We see that no tracks modelled with $\mathrm{AML_{GR}}$ evolve to match either ZTFJ1637+49 or Gaia14aae. For \textcolor{black}{Models 1, 3, and 5} we see that neither the orbital period nor the radius of the donor increase enough during the expansion phase of the system. In the ($P_\mathrm{orb},\,\Dot{M}_\mathrm{He}$) plane, the mass-transfer rate does not increase enough to match that of Gaia14aae. However, for \textcolor{black}{Models 6 and 7} we see that there is not much difference in these parameters. To investigate this, we plot the time-scale of $\mathrm{AML_{GR}}$ ($\tau_\mathrm{GR}$), given by $\tau_\mathrm{GR} = J/\Dot{J}_\mathrm{GR}$ and the time-scale of $\mathrm{AML_{DD}}$ ($\tau_\mathrm{DD}$), given by $\tau_\mathrm{DD} = J/\Dot{J}_\mathrm{DD}$. In order to understand how the two dynamos in the DD model contribute to the evolution of the system, we also plot the time-scale of AML solely due to the boundary layer dynamo, $\mathrm{AML_{bl}}$ ($\tau_\mathrm{bl}$) given by $\tau_\mathrm{bl} = J/\Dot{J}_\mathrm{bl}$, and the time-scale of AML solely due to the convective dynamo, $\mathrm{AML_{conv}}$ ($\tau_\mathrm{conv}$) given by $\tau_\mathrm{conv} = J/\Dot{J}_\mathrm{conv}$. We note that the combined effect of the two dynamos is not simply additive, as can be seen from equations (\ref{eq:jdotbyjdd}) to (\ref{eq:bptot}). Our results are shown in Fig.~\ref{fig:dynamo}, with subplots titled by the \textcolor{black}{model numbers} in Fig.~\ref{fig:prmmdot_gr}. We see that \textcolor{black}{Models 3 and 5} have $\tau_\mathrm{GR}\approx\tau_\mathrm{DD}$ for the entire evolution, with $\tau_\mathrm{DD}\lesssim\tau_\mathrm{GR}$ at later times. This explains why the systems evolved with $\mathrm{AML_{GR}}\,+\,\mathrm{AMl_{DD}}$ have longer orbital periods and donor radii for the same donor mass. So at later times, owing to $\mathrm{AML_{DD}}$, the time-scale of mass-loss does not increase significantly. This keeps the Kelvin-Helmholtz time-scale ($\tau_\mathrm{KH}$) larger than the mass-loss time-scale ($\tau_\mathrm{ML}$) or, in other words, the system continues to respond adiabatically to mass loss. This is in contrast to the results of \cite{Deloye2007} and \cite{Yungelson2008} in which, after the period minimum, the adiabatic expansion phase of the system ceases once $\tau_\mathrm{KH}\lesssim \tau_\mathrm{ML}$. This has been hinted at by \cite{2022MNRAS.512.5440V} who argue that for some reason the orbital periods of AM CVn systems probably do not shrink when they are long.

We also see that $\tau_\mathrm{DD}$ is dominated by $\tau_\mathrm{conv}$ for most of the evolution, with $\tau_\mathrm{conv}$ becoming increasingly dominant as the donors tend towards full convection. \textcolor{black}{This behaviour is an extension of that in canonical CVs where, at the upper end of the period gap ($P_\mathrm{orb}\approx 3\,\mathrm{hr}$), $\tau_\mathrm{bl}$ is dominant and once RLOF recommences below the period gap ($P_\mathrm{orb}\approx 2\,\mathrm{hr}$) $\tau_\mathrm{GR}$ drives the evolution with increasing contribution by $\tau_\mathrm{conv}$.} However, for \textcolor{black}{Models 6 and 7}, we see that $\tau_\mathrm{GR}\lesssim\tau_\mathrm{DD}$ for most of the evolution. This can be explained by the fact that $\tau_\mathrm{GR}\propto a^4$ whereas $\tau_\mathrm{DD}\propto a^2$ and so $\tau_\mathrm{GR}$ dominates at $P_\mathrm{orb}\approx 10\,\mathrm{min}$. The evolution of these systems can be well modelled with just $\mathrm{AML_{GR}}$.  For \textcolor{black}{Model 6} we see that $\tau_\mathrm{bl}\approx\tau_\mathrm{conv}$, with $\tau_\mathrm{bl}\lesssim\tau_\mathrm{conv}$ for \textcolor{black}{Model 7}. This is explained by equations (18) to (25) of ST in which $\tau_\mathrm{bl}\propto \rho_\mathrm{B}^{1/4}$, where $\rho_\mathrm{B}$ is the density at the boundary between the core and the envelope. The density $\rho_\mathrm{B}$ increases in more degenerate configurations causing the boundary layer dynamo to dominate $\tau_\mathrm{DD}$. 
 
\section{Conclusion}
\label{sec:conc}

We have revisited the He-star channel of AM CVn formation to explain Gaia14aae and ZTF1637J+49, peculiar AM CVn systems which seemingly challenged the current understanding of their formation. \textcolor{black}{We argue that semi-degenerate, He-rich donors with masses between $0.1$ and $0.3\,M_\odot$ plus WD binaries which are able to explain these systems can emerge from a common envelope phase if either sources of energy other than the orbital energy are used to eject the common envelope, or a different formalism of common envelope evolution is implemented.} We model this by simply setting a maximum allowed common envelope efficiency parameter larger than unity. Using a set of He-star plus WD binaries which emerge from this common envelope phase as our initial models, we track the detailed evolution of orbital parameters and surface abundances with the Cambridge stellar evolution code (\textsc{STARS}), when the evolution is driven, not only by the loss of angular momentum in gravitational waves, but also by magnetic braking owing to an $\alpha-\Omega$ dynamo mechanism. Such a model leads to the period gap and the period minimum spike in the orbital period distribution of cataclysmic variables. Using the same physically motivated angular momentum loss formalism as \cite{2022MNRAS.513.4169S} in addition to gravitational radiation, we show that our modelled trajectories match well with the observed properties and abundance profiles of Gaia14aae and ZTF1637J+49. We also highlight the importance of including the additional angular momentum loss mechanism in our analysis by showing that our selected trajectories fail to reproduce the observed properties of Gaia14aae and ZTF1637J+49 when the evolution is driven solely by angular momentum loss in gravitational radiation.

There are a few areas of improvement to be made to our analysis. Most important is modelling the common envelope evolution phase which forms the He-star plus WD binary. Our analysis in Section~\ref{sec:cee} is kept quite straightforward, because this phase is poorly understood. Detailed simulations of the event and knowledge of the interplay between the additional energy sources that may be used by the system to eject the common envelope will help to improve the situation, especially when there is growing evidence of systems with seemingly unphysical common envelope ejection efficiency parameters (e.g. \citealt{ElBadry2022}). This in turn will either strengthen or weaken the claim that our semi-degenerate He-rich donors emerge from the common evolution phase in a close orbit with a WD. Same improvement should be made to the additional angular momentum loss mechanism arising as a consequence of the DD model. This mechanism was implemented with the help of three free parameters by \cite{2022MNRAS.513.4169S} and fitted in order to explain the properties of cataclysmic variables. Although the formalism of the DD model is physical, the free parameters themselves are ad hoc and may need to be modified if we extend our analysis to evolved donors. This, with more accurate measurements, can address the differences between the mass-transfer rates in our models and observations. Finally, there is a need for a detailed population synthesis study of all the AM CVn evolution channels, with new results on the relative likelihood of AM CVn formation (e.g. \citealt{Shen2015}). This will provide clearer insights into the relative importance of the three proposed AM CVn formation channels.

\section*{Acknowledgements}
\textcolor{black}{The authors thank the referee, Prof. Zhanwen Han, for his detailed review of the manuscript. His helpful comments and suggestions greatly enhanced the overall structure and content of this work.} AS thanks the Gates Cambridge Trust for his scholarship. HG acknowledges support from NSFC (grant No. 12173081), the key research program of frontier sciences, CAS, No. ZDBS-LY-7005, and Yunnan Fundamental Research Projects (grant No. 202101AV070001). CAT thanks Churchill
College for his fellowship. 

\section*{Data availability}
No new data were generated or analysed in support of this research. Any numerical codes and related data generated during the work will be made available whenever required by the readers.

\bibliographystyle{mnras}
\bibliography{mnras_template} 

\bsp	
\label{lastpage}
\end{document}